# NOISE IS ALL YOU NEED: RETHINKING THE VALUE OF NOISE ON SEISMIC DENOISING VIA DIFFUSION MODELS

Donglin Zhu[1], Peiyao Li[1], and Ge Jin[1]

1. Department of Geophysics, Colorado School of Mines, Colorado, USA. E-mail: dzhu@mines.edu, lipeiyao@mines.edu, gjin@mines.edu

ABSTRACT

We introduce SeisDiff-denoNIA, a diffusion-based seismic denoising framework that trains directly on noise extracted from field data, eliminating the dependence on synthetic datasets. Unlike conventional denoising methods that require clean signal labels, our approach uses field noise recorded as training targets, enabling the diffusion model to explicitly learn the true noise distribution. We further demonstrate the framework on a field DAS-VSP survey, where the model effectively suppresses multiple types of noise, including production, instrument, and environment related noise, while preserving key seismic events. By denoising shot gathers with SeisDiff-denoNIA prior to migration, the resulting images exhibit improved event continuity and enhanced reflection visibility without artificial artifacts. In addition, synthetic experiments show that the method substantially outperforms traditional signal-based diffusion models under low-SNR conditions. These results suggest that explicitly modeling noise is not only viable but advantageous for a broad class of seismic denoising tasks, particularly in challenging field environments.

# INTRODUCTION

Seismic data acquisition is inherently susceptible to various types of noise, which can obscure critical subsurface information and impede accurate interpretation. These noise sources include environmental factors, equipment limitations, and human activities, all contributing to the complexity of seismic signals. Effective denoising is, therefore, essential to enhance the signal-to-noise ratio (SNR) and ensure reliable analysis of seismic data.

Traditional denoising techniques, such as filtering and transform-based methods, often rely on assumptions about the noise characteristics or the signal's sparsity in a particular domain. The f-x filtering utilizes the predictability of seismic events in the frequency-space domain to suppress noise (Abma and Claerbout, 1995). Wavelet transform filtering decomposes seismic signals into different frequency components, allowing for targeted noise attenuation (Deighan and Watts, 1997). Empirical mode decomposition decomposes signals into intrinsic mode functions to isolate and remove noise components (Bekara and van der Baan, 2009). While these traditional filtering methods have proven effective, they often rely on assumptions about noise characteristics and may struggle in low SNR conditions. This limitation has prompted the exploration of more adaptive and data-driven approaches to seismic denoising.

In recent years, machine learning (ML) has emerged as a powerful tool in seismic data processing, offering adaptive algorithms capable of learning complex patterns directly from data. Supervised learning approaches, in particular, have shown promise in denoising applications by training models to distinguish between signal and noise (Saad and Chen, 2020; Yang et al., 2023; Dong et al., 2022). However, these methods typically require large datasets with clean signal labels, which are often unavailable in practical scenarios. Consequently, synthetic data is frequently employed for training purposes. While synthetic datasets can be meticulously crafted to represent ideal conditions, they often fail to capture the full complexity and variability of real field data, leading to models that may not generalize well to actual seismic recordings. Some unsupervised learning methods (Liu et al., 2023; Wang et al., 2023; Konietzny et al., 2024; Luiken et al., 2024) try to overcome these problems, but they may struggle to achieve the same level of accuracy and effectiveness as supervised approaches, especially when noise characteristics closely resemble those of the signal, leading to potential signal loss or incomplete noise attenuation.

Recent advancements in generative models, particularly diffusion models for seismic denoising, have demonstrated remarkable capabilities in noise attenuation and signal reconstruction. Durall et al. (2023) applied the denoising diffusion probabilistic model (DDPM) (Ho et al., 2020) for multiple suppression in seismic data. Similarly, diffusion models have been employed to mitigate coupling noise and fading noise in DAS-VSP data (Zhu et al., 2023) and to attenuate ground-roll noise (Li et al., 2024). More recently, Zhu et al. (2025) proposed a diffusion model-based workflow for tube wave attenuation. While these diffusion model based studies show promising results compared to current traditional methods and neural networks, they still largely rely on synthetic seismic data with noise as labels or targets, a persistent limitation in supervised learning-based seismic denoising. Synthetic seismic data and artificially generated noise are often overly simplistic and fail to capture the complexities of field recordings. As a result, models trained on synthetic or semi-synthetic datasets tend to suffer from inaccuracies and poor generalization when applied to field data.

To overcome these challenges, we revisit diffusion model-based approaches for seismic denoising and propose SeisDiff-denoNIA (Figure 1), a fully field noise data-driven workflow that eliminates dependence on synthetic data. Our approach re-evaluates the role of noise in field recordings and leverages a diffusion model that directly learns and reconstructs noise from real seismic data. In this paper, we first present the methodology, outlining the design of our diffusion model and the use of field noise as training targets. Next, we describe the experimental setup and evaluate the model's performance on both field data and quantitative tests using synthetic data. Finally, we discuss the effectiveness of the proposed approach and summarize our key findings.

## METHODOLOGY

Diffusion models such as DDPM learn the underlying data distribution and reconstruct signals by progressively denoising sampled noise. We adopt SeisDiff-deno workflow from Zhu et al. (2025), and evolve it from relying on synthetic/semi-synthetic seismic to full field data driven mode. Our key innovation is leveraging seismic noise extracted directly from field data as the training target, rather than relying on synthetic seismic or processed field seismic data by traditional methods. By explicitly learning the statistical characteristics of field noise distributions, independent of seismic signals, the Conditional Diffusion Probabilistic Model ensures robust generalization across various field seismic datasets. Additionally, to accelerate inference while maintaining performance, we incorporate the Diffusion Probabilistic Model solver (DPM-solver) (Lu et al., 2022a, 2022b; Zheng et al., 2023) during the sampling process.

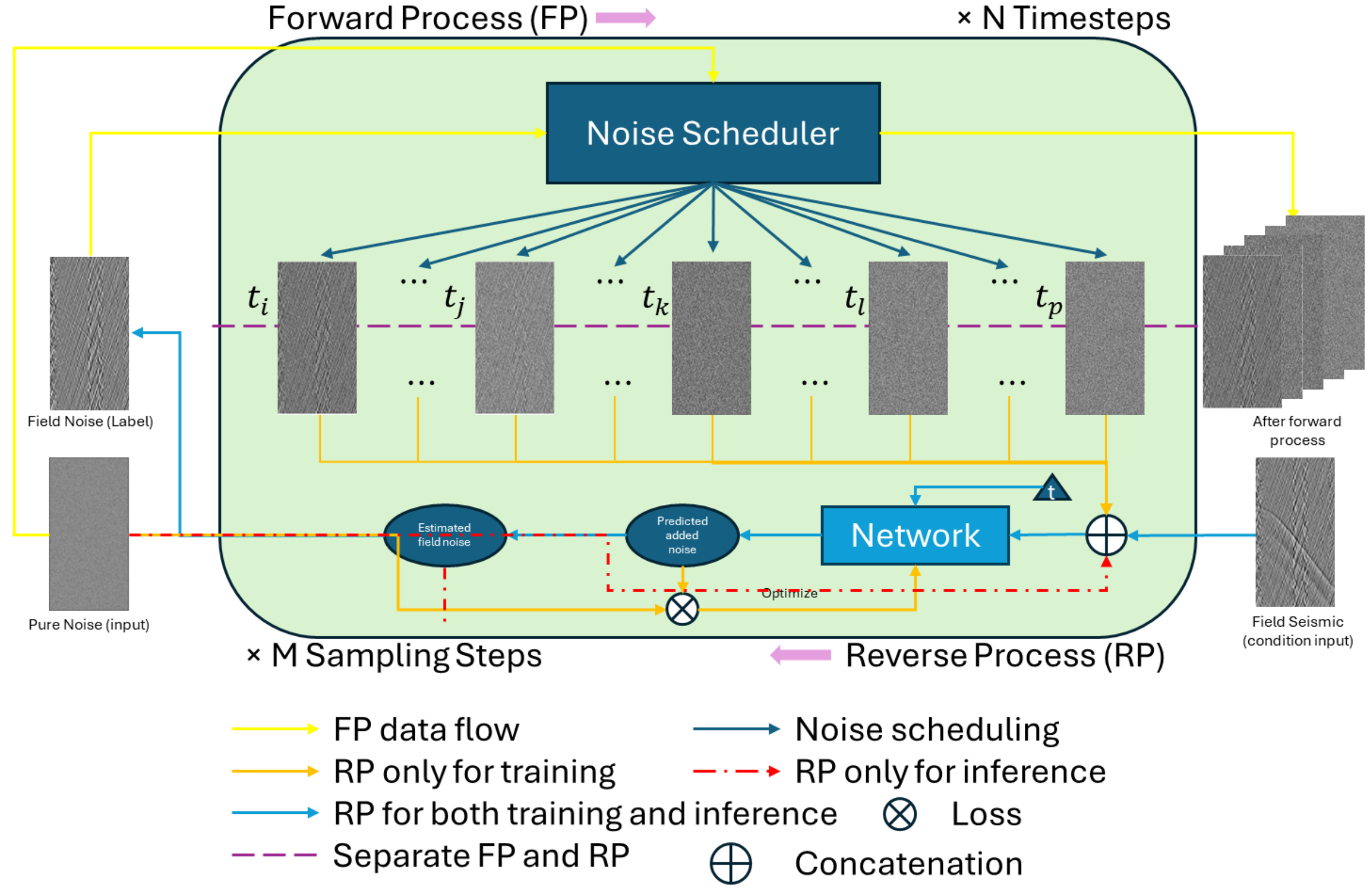

Figure 1. SeisDiff-denoNIA workflow. The FP follows conditional DDPM, and the RP incorporates DOM-solver. The figure is modified from Zhu et al. (2025)

### Training data preparation

To prepare the training dataset, real seismic noise was carefully extracted directly from field seismic gathers by isolating segments captured before the first arrival of seismic waves. This approach ensures the quality of noise samples since these segments inherently lack seismic signal components. Mathematically, this extraction process is expressed as:

$$n_{field}(t) = s_{field}(t), for\ t < t_{first_arrival}, \tag{1}$$

where $n_{field}(t)$ is the extracted seismic noise, $s_{field}(t)$ represents the field seismic data, and $t_{first_arrival}$ denotes the time of the first arrival of seismic data. This ensures the extracted noise accurately captures the field-specific noise characteristics and distribution without contamination from primary reflections or other seismic events. After obtaining the noise label, we introduce controlled perturbations, which could disturb noise distribution or pattern, to the noise label to construct the training samples. We further examine how different types of perturbations influence the training performance in the discussion section. An example of training dataset pair is shown in Figure 2.

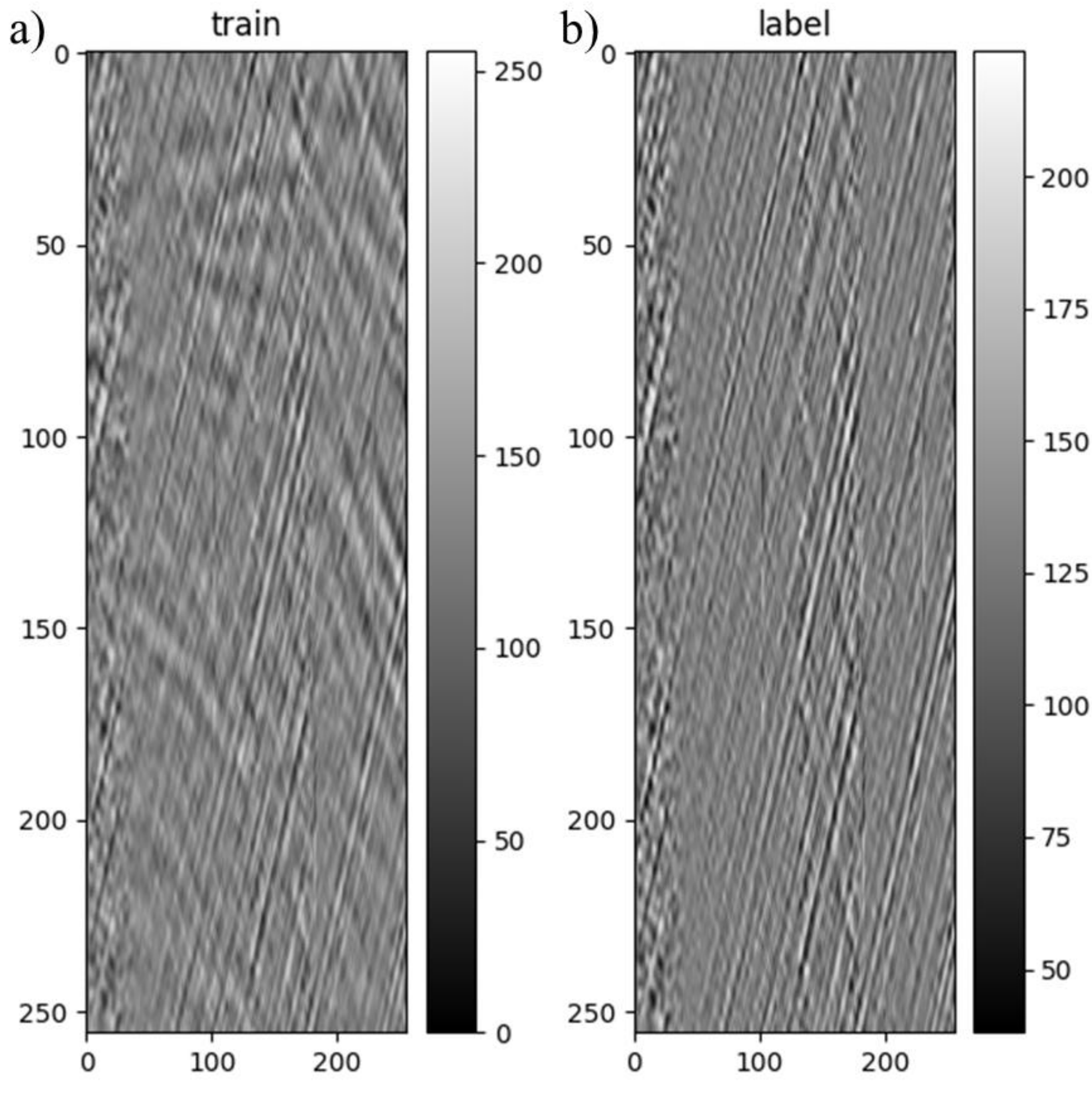

Figure 2. Example of training data pair. a) noise with perturbation as train data, b) extracted noise as label data.

**Diffusion Denoising Probabilistic Model**

Unlike traditional diffusion models application for denoising tasks which target seismic signals, our approach explicitly models the noise distribution itself. During training, the extracted seismic noise serves as the primary target. To ensure compatibility with the diffusion process, this real noise is subjected to additional Gaussian noise increments, resulting in purely Gaussian noise at the last diffusion timestep.

The diffusion probabilistic model comprises a forward and reverse diffusion processes. In the forward diffusion stage, Gaussian noise is iteratively added to following a predefined schedule, defined as:

$$q(x_t|x_{t-1}) = N(x_t; \sqrt{1-\beta_t}x_{t-1}, \beta_t\mathbf{I}), \tag{2}$$

$$x_t = \sqrt{\bar{\alpha}_t}x_0 + \sqrt{1-\bar{\alpha}_t}\epsilon, \tag{3}$$

where $q(x_t|x_{t-1})$ represents progressively noisier versions at timestep t, $\beta_t$ is the factor controlling the added noise, and $\bar{\alpha}_t$ denotes the products of $1-\beta_t$ .

In the reverse diffusion process, the neural network iteratively removes this Gaussian noise, conditioned on noisy seismic data that includes the real extracted noise plus other potential signals. Specifically, the conditional probability distribution is modeled as:

$$p_\theta(x_{t-1}|x_t, c) = N(x_{t-1}; \mu_\theta(x_t, c, t), \Sigma_\theta(x_t, c, t)\mathbf{I}), \tag{4}$$

with $\theta$ predicted by a neural network, $\mu_\theta$ is the mean, $\Sigma_\theta$ represents variance, and the conditional input $c$ comprises the extracted noisy seismic.

## Network Architecture

The employed neural network follows a modified U-Net architecture with attention blocks (Zhu et al, 2025) specifically designed for seismic denoising tasks. It incorporates residual blocks to facilitate efficient gradient propagation and attention mechanisms to effectively capture contextual relationships across seismic traces. Residual blocks enable deeper networks by alleviating gradient vanishing issues, while attention blocks enable the model to selectively focus on significant spatial-temporal regions, thus enhancing noise reconstruction accuracy.

The conditional embedding is integrated into the U-Net structure, where noisy seismic data serves as the conditional input to guide the denoising process. Each level of the U-Net comprises convolutional layers followed by attention modules and residual connections, which collectively help the network to robustly learn complex noise patterns. The full parameters of network is shown in Table 1.

Tabel 1. Neural network parameters

| Name | Parameters |
| --- | --- |
| Input Channel | 2 |
| Output Channel | 1 |
| Number of Residul Block | 2 |
| Attention Resolution | 16 |
| Base Filter Number | 128 |
| Convolution Kernel Size | 3 |
| Convolution Stride | 1 |
| Convolution Padding | 1 |
| Multiple of Filter Number | [1,1,2,2,4,4] |
| Upsampling Scale Factor | 2 |
| Normalize mode | Group Norm |
| Timestep embedding | Sinusoidal Embeddings |
| Condition embedding | Concatenate |

**Accelerated Sampling via DPM-Solver**

To significantly enhance computational efficiency without sacrificing denoising quality, we incorporated the DPM-solver during the sampling stage. The DPM-solver employs numerical techniques to solve the ordinary differential equation (ODE) representation of the diffusion process, allowing accelerated inference by reducing the required sampling steps. The ODE formulation for the diffusion process is given by:

$$\frac{dx_t}{dt} = -\frac{1}{2}\beta_t\left[x_t + 2\nabla_{x_t}\log p_\theta(x_t|c)\right], \tag{5}$$

where $\nabla_{x_t} \log p_\theta(x_t|c)$ represents the score function, explicitly related to the neural network's predicted noise $\epsilon_\theta(x_t, c, t_i)$ by:

$$\nabla_{x_t} \log p_\theta(x_t|c) = -\frac{\epsilon_\theta(x_t,c,t_i)}{\sqrt{\beta_t}}, \tag{6}$$

This explicit relationship connects the DDPM and DPM-solver formulations, ensuring consistent notation and clear linkage between the two processes. The DPM-solver numerically integrates the ODE efficiently, significantly reducing the number of required sampling steps by:

$$x_{t-\Delta t} = x_t + \sum_{i=1}^{k} a_i \epsilon_\theta(x_t, c, t_i), \tag{7}$$

where coefficients $a_i$ depend on the numerical solver used, and $\epsilon_\theta(x_t, c, t_i)$ denotes the neural network-based denoising function.

**Implementation Details**

The diffusion model was implemented using PyTorch, trained on Google Colab A100 GPU with 80GB memory. Each training epoch consisted of randomly sampled noise segments from field seismic gathers captured before seismic arrivals, ensuring authenticity of noise distribution. The training dataset was further augmented through data augmentation techniques, including random cropping and normalization, to improve generalization.

Our training approach involves minimizing a mean squared error (MSE) loss function defined as the difference between the predicted noise and the actual extracted seismic noise:

$$L_t = E_{t,x_0,c,\theta,\epsilon} \|\epsilon - \epsilon_\theta(x_t, t, c)\|^2, \tag{8}$$

where $\epsilon$ is the real noise introduced at timestep $t$, and $\epsilon_\theta(x_t, t, c)$ represents the predicted noise from the neural network. Model optimization utilized the AdamW optimizer with a learning rate set to $2 \times 10^{-4}$, and training continued until convergence was observed, typically after several hundred epochs. The trained model was validated against independent field datasets to ensure robustness and generalization capabilities. The inference is conducted on a local NVIDIA RTX A4000 with 16GB memory.

In summary, the proposed workflow, leveraging real seismic noise extracted directly from field recordings and accelerated inference via the DPM-solver, provides an effective and computationally efficient solution for seismic data denoising tasks. The timesteps set to 1000 in forward process, and 20 to reverse process. This approach enhances model robustness by explicitly learning noise distributions, ensuring superior generalization performance in field seismic data applications.

## APPLICATION

We validate the proposed SeisDiff-denoNIA framework through experiments on both field and synthetic datasets. For real-world evaluation, we apply the model to DAS-VSP data from the offshore Texas, which is heavily contaminated by different types of noise. Additionally, synthetic tests using SEAM Phase I elastic VSP data are conducted to quantify performance under controlled SNR conditions. We use the signal-to-noise ratio (SNR) to evaluate the denoising results for testing quantitively, the cross-correlation value (CC) to measure the structural similarity, and the relative root mean square error (RMSE) to evaluate the amplitude discrepancy between the denoised image and the ground truth image if the ground truth is known. The SNR is defined as:

$$SNR = 10log_{10}\left(\frac{P_{signal}}{P_{noise}}\right), \tag{9}$$

where $P_{signal}$ is the power spectrum of the signal, and $P_{noise}$ is the power spectrum of the noise computed from $n_{field}(t)$. The cross-correlation value can be defined by:

$$CC(x,\bar{x}) = \frac{\sum_{i=1}^{N}(x_i-\bar{x})(y_i-\bar{y})}{\sqrt{\sum_{i=1}^{N}(x_i-\bar{x})^2 \sum_{i=1}^{N}(y_i-\bar{y})^2}}, \tag{10}$$

where $\bar{x}$ $and$ $\bar{y}$ are the mean value of clean data and denoised data, respectively. The RMSE can be defined:

$$RMSE(x,\bar{x}) = \sqrt[2]{(\mu_x - \mu_{\bar{x}})^2} \tag{11}$$

where $\mu_x$ and $\mu_{\bar{x}}$ are the mean values of clean data and denoised data, respectively.

**Field test**

The field data is a 4D DAS-VSP dataset from a deepwater field. The VSP data are recorded by the multi-mode fiber optical cables located in two wells, which are both active injector wells, from 2015 to 2018 (Zwartjes et al., 2018). Some of the shots experience strong tube wave contamination due to fluid flow in the wells. Others are tube wave free when the operators shut in the well but are still contaminated with strong noise. We extract field noise from 800 shots before the first arrivals to generate a training dataset, and 16000 patches in size of 256 by 256 are captured from shot gathers for training. We then apply the trained model to several DAS-VSP shot lines with different noise contaminations.

The proposed method specifically targets the noise itself while removing all other information, and the clean shot could be obtained from subtracting prediction noise from the noisy shot. Due to different types of noise existing in the dataset, we visualize them on Figure 3 and Figure 4, respectively. Figure 3 presents a shot primarily dominated by tube wave noise. After applying SeisDiff-denoNIA, the tube wave energy is thoroughly suppressed, while underlying reflection signals are preserved—even in regions heavily blended with the noise. Additionally, minor noise types such as DAS fading and cable coupling noise are also attenuated, showing the model's capacity to generalize across complex noise compositions. Figure 5 shows the time window focused on the first arrival (1.6-3.2s) in green box from Figure 3b. Within this window, the noise is effectively suppressed, and the PS-converted wave is successfully recovered, even though it is blended with the tube wave in the noisy shot. To further illustrate the capabilities of dealing with different wavefield, we investigated other shots where the noise type is quite different. Figure 5a shows a shot without tube wave but still contaminated strongly. The model successfully retains the seismic signal even in the presence of strong noise. This capability is critical for subsequent imaging processes. To further assess signal recovery from relative deep recording time, we analyze zoomed-in time windows from 6–8s, highlighted in red box in Figure 4b. The corresponding zoomed denoising results are shown in Figure 6, demonstrating the model's ability to preserve signal continuity at depth. These results suggest that SeisDiff-denoNIA not only effectively suppresses strong complex noise but also preserves weak signals vital for imaging and interpretation.

Following the denoising stage, we apply Kirchhoff prestack depth migration workflow (Figure 7) to two datasets for comparison: (1) shot gathers denoised using the proposed diffusion model (Figure 8b) and (2) shot gathers processed using conventional FK filtering (Figure 8c) only. This parallel design allows us to directly evaluate the impact of diffusion-based denoising on subsequent imaging quality.

For both workflows, source and receiver traveltime fields are computed by solving the Eikonal equation using the Fast Sweeping Method (FSM) based on the input velocity model,

$$|\nabla T(x)^2| = \frac{1}{v^2(x)}, \quad (12)$$

where T(x) denotes the first-arrival traveltime and v(x) is the spatially varying velocity. These traveltimes define the kinematics for the Kirchhoff migration operator. Prior to migration, the recorded wavefield is decomposed into upgoing and downgoing components using a frequency-slowness $(\omega, p)$ $\tau - p$ transform along the receiver-depth axis $z$. Let $U(\omega, z)$ denote the temporal Fourier transform of the data recorded at depth $z$ (with $z$ positive downward). The forward $\tau - p$ transform is

$$S(\omega, p) = \int U(\omega, z) e^{-i\omega p z} dz, \quad (13)$$

where $\omega$ is angular frequency, and $p$ is the apparent vertical slowness associated with a plane wave component such that the moveout satisfies $t(z) \approx pz$. In practice, the integral is implemented as a discrete summation over receiver depths.

Directional separation is achieved by filtering in $p$. Under the positive-downward depth convention used here, wave modes are distinguished by the sign of the depth moveout:

- Downgoing arrivals satisfy $\frac{dt}{dz} < 0$, corresponding to negative slowness $p < 0$.
- Upgoing arrivals satisfy $\frac{dt}{dz} > 0$, corresponding to positive slowness $p > 0$.

Thus, the upgoing and downgoing components in the $\omega - p$ domain is obtained by applying directional masks:

$$S_u(\omega, p) = S(\omega, p) 1_{(p>0)}, \; S_d(\omega, p) = S(\omega, p) 1_{(p<0)}, \quad (14)$$

The separated components are then mapped back to the receiver-depth domain via inverse $\tau - p$ transform,

$$U_u(\omega, z) = \frac{1}{N_p} \sum_p S_u(\omega, p) e^{+i\omega p z} \quad (15)$$

Finally, the upgoing wavefield is transformed back to the time domain by inverse Fourier transform,

$$U_u(t, z) = \mathrm{F}^{-1}\{U_u(\omega, z)\} \quad (16)$$

Kirchhoff migration is then performed by summing amplitudes along two-way traveltime surfaces,

$$I(x) = \sum_r w(x, r) u_{u,r}(T_s(x) + T_r(x)), \quad (17)$$

where $u_r(t)$ is the upgoing wavefield at receiver r, $T_s$ and $T_r$ are the source- and receiver-to-image traveltimes, respectively, and $w(x, r)$ denotes the geometric-spreading weight. Aperture control is applied by restricting contributions to rays whose opening angle falls within a prescribed range (80 degrees in this case). As a final post-processing step applied consistently to both imaging results, a horizontal Laplacian filter is applied in the depth domain which enhances reflector sharpness by emphasizing lateral discontinuities and suppressing residual low-wavenumber blurring, following a common practice in seismic imaging to improve the visual interpretability of reflectivity structures.

Because the fiber-optic cable follows a 3D well trajectory while our imaging algorithm targeted for 2D slice, we first selected the near-vertical portion of the well that contains the DAS receivers. The receiver coordinates $(x, y)$ were then used to fit a representative receiver line on the surface shot map (Figure 9). Shots falling within a small tolerance (< 3m) of this fitted line were retained. Finally, a 2D velocity-model section was extracted from the 3D velocity cube along the plane defined by the receiver line and the selected shot line for use in the 2D migration.

For the diffusion-denoised data, the migrated images exhibit laterally coherent, well-focused, and laterally flattened reflectors, indicating that the diffusion model effectively suppresses coherent tube wave while preserving true reflection energy (Figure 10a). The improved signal continuity directly translates into enhanced reflector sharpness and reduced imaging ambiguity at depth.

In contrast, when the same imaging workflow is applied to FK-only filtered shot gathers, the resulting migrated section displays severe migration artifacts (Figure 10b). These artifacts manifest as spurious energy both along reflector interfaces and throughout the imaging domain, reflecting residual coherent and incoherent noise that is not adequately removed by FK filtering alone. The presence of these artifacts severely degrades reflector continuity and compromises structural interpretation.

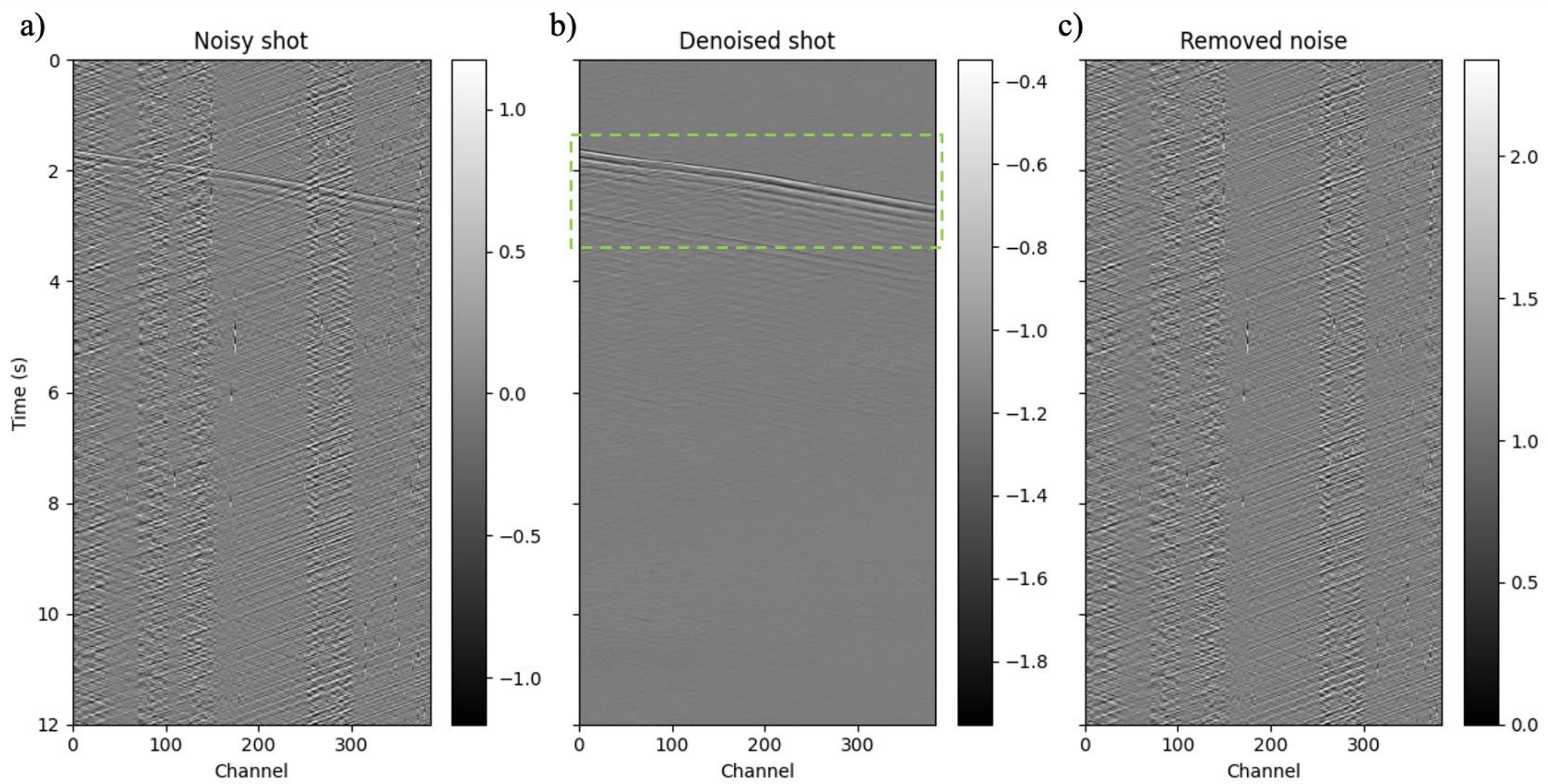

Figure 3. Field test with tube wave contaminated shot, a) noisy shot, b) denoised result, and c) removed noise.

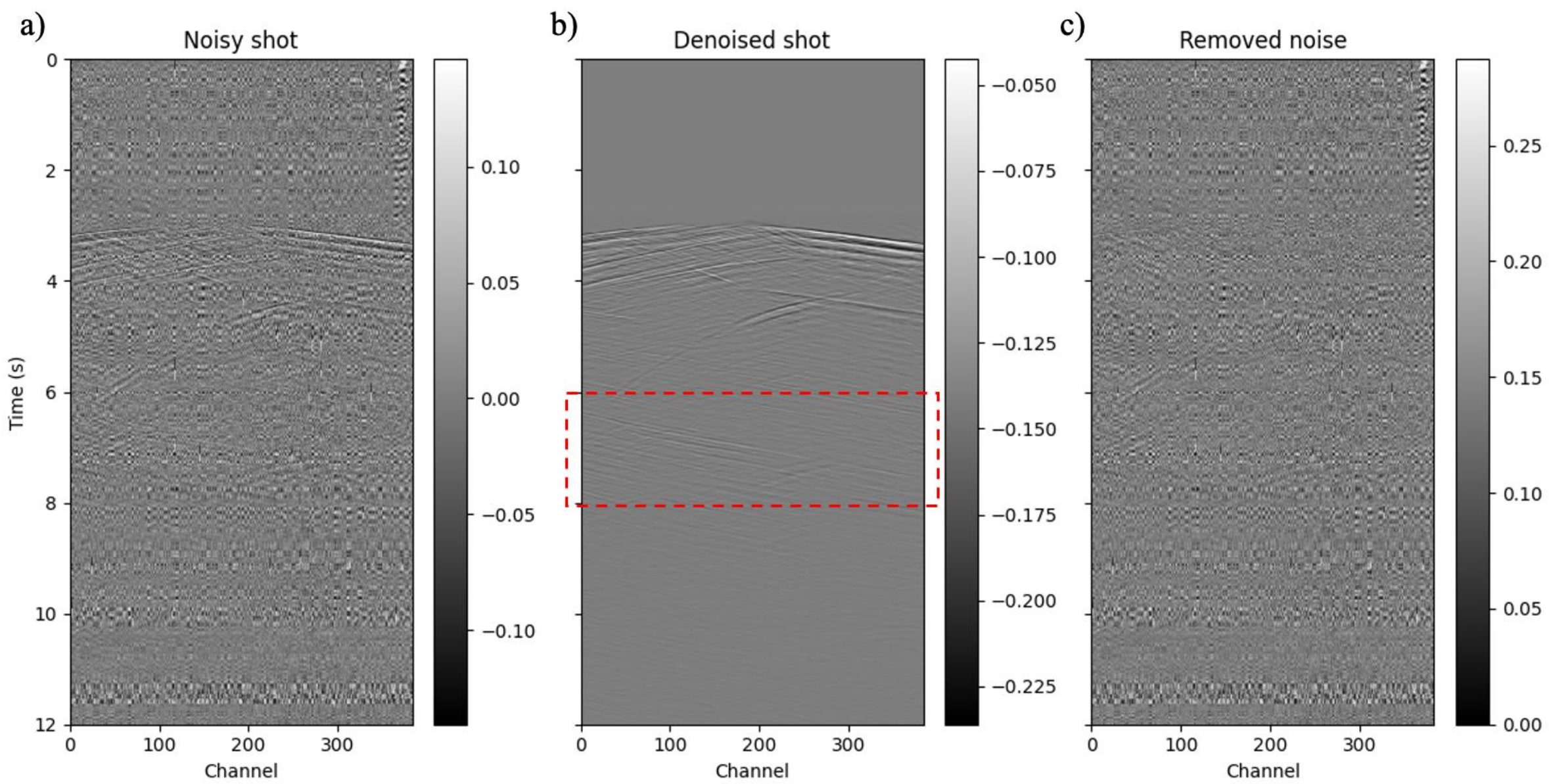

Figure 4. Field test with tube wave free shot. a) Field DAS-VSP record, b) denoising results, and c) removed noise.

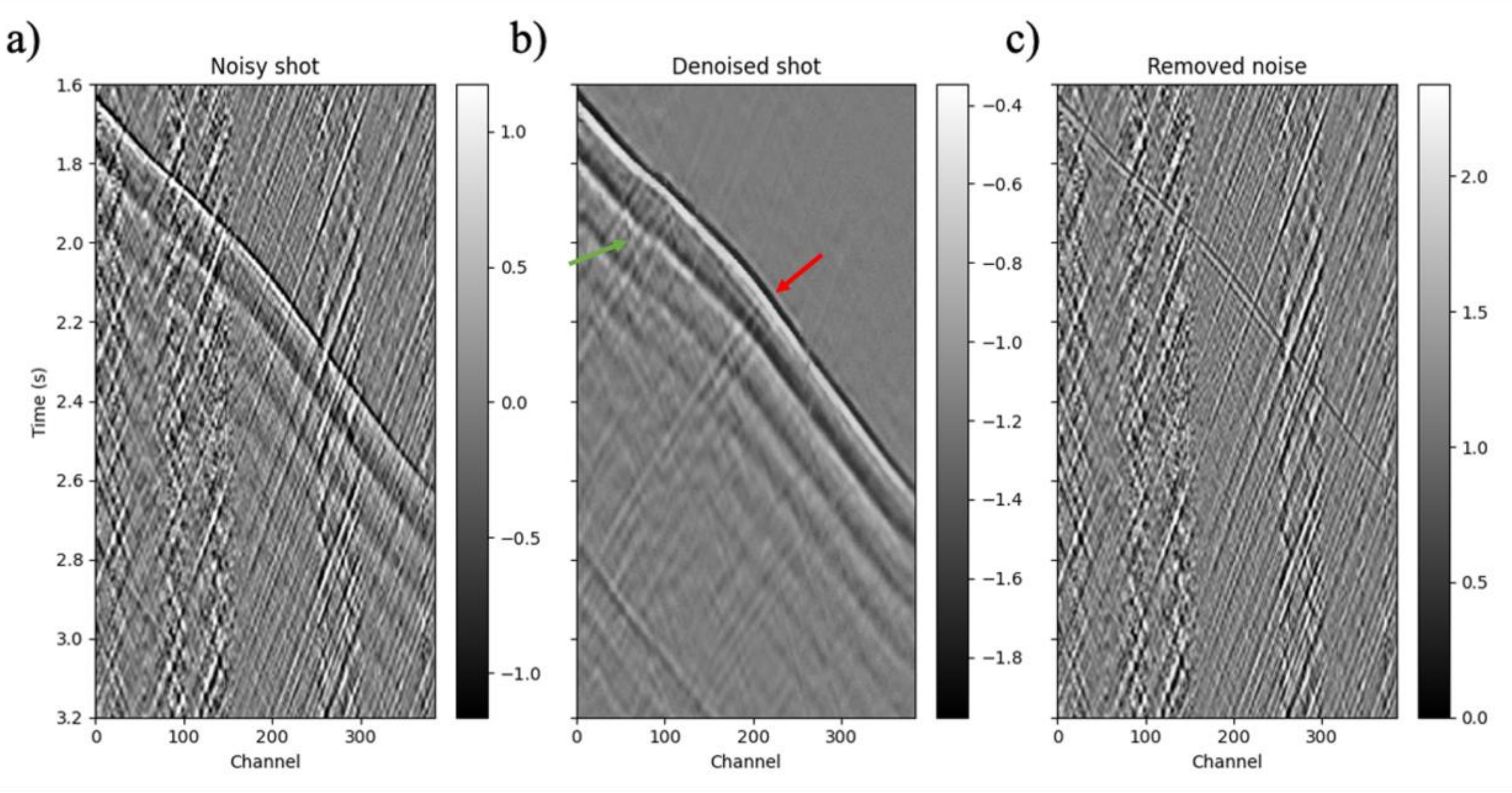

Figure 5. Zoomed in part of the selected time (1.6-3.2s) windows box (green box) on Figure 3. a) noisy shot section, b) denoised result, c) removed noise. The red arrow indicates to the direct P wave, the green arrow indicates to the PS converted wave.

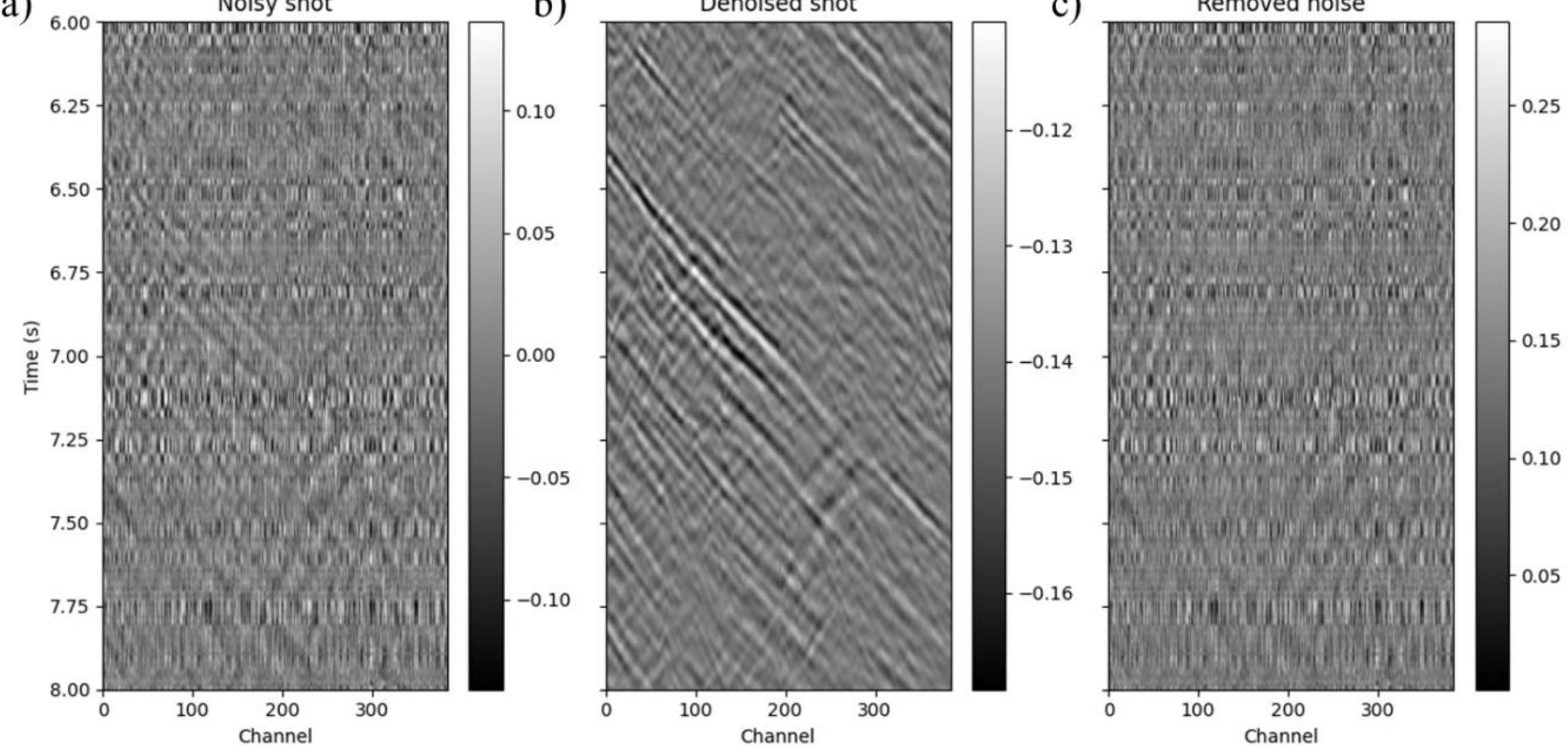

Figure 6. Zoomed in part of the selected time (6-8s) windows box (red box) on Figure 4. a) noisy shot section, b) denoised result, c) removed noise.

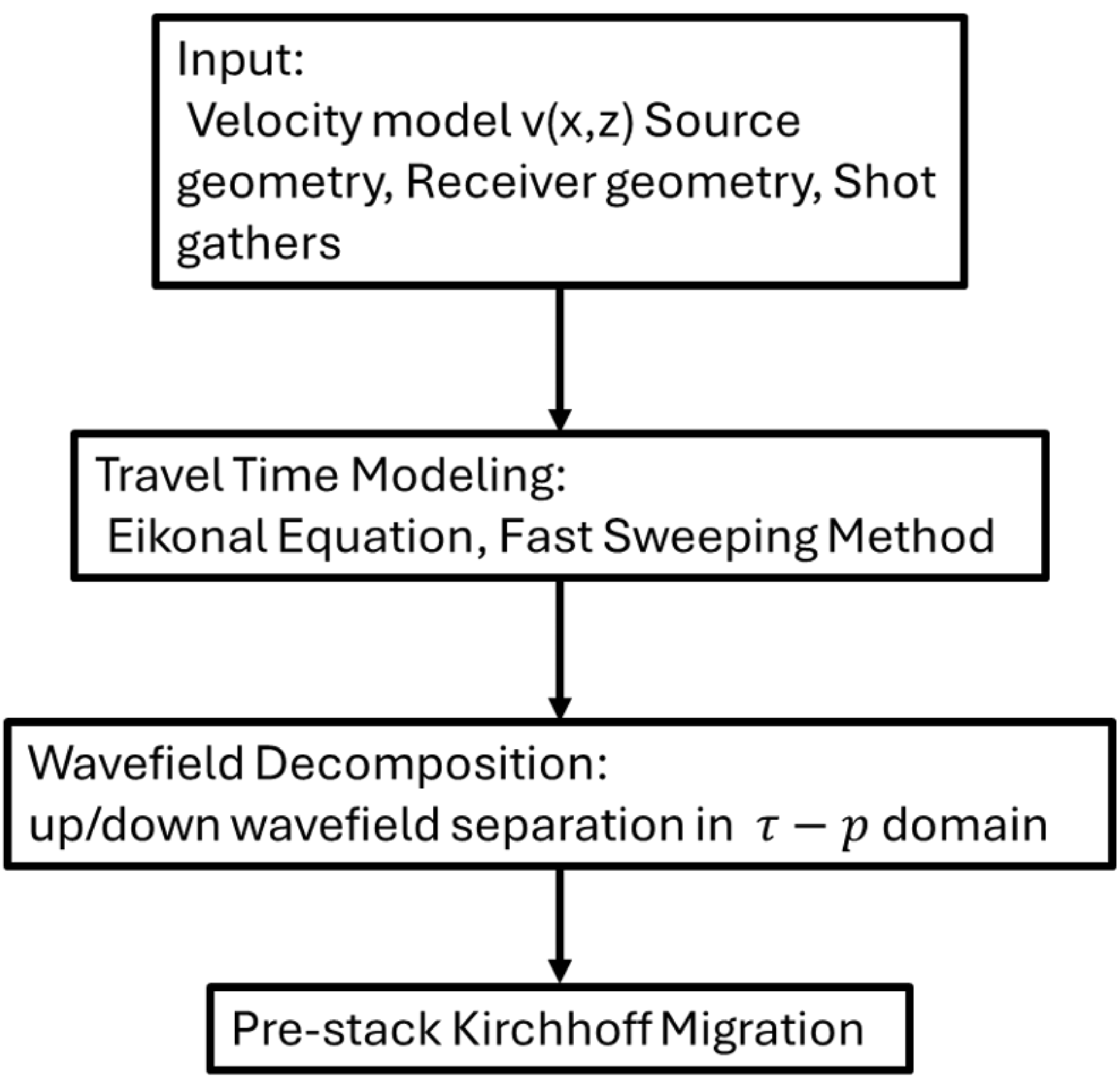

Figure 7. Migration worfklow for VSP data

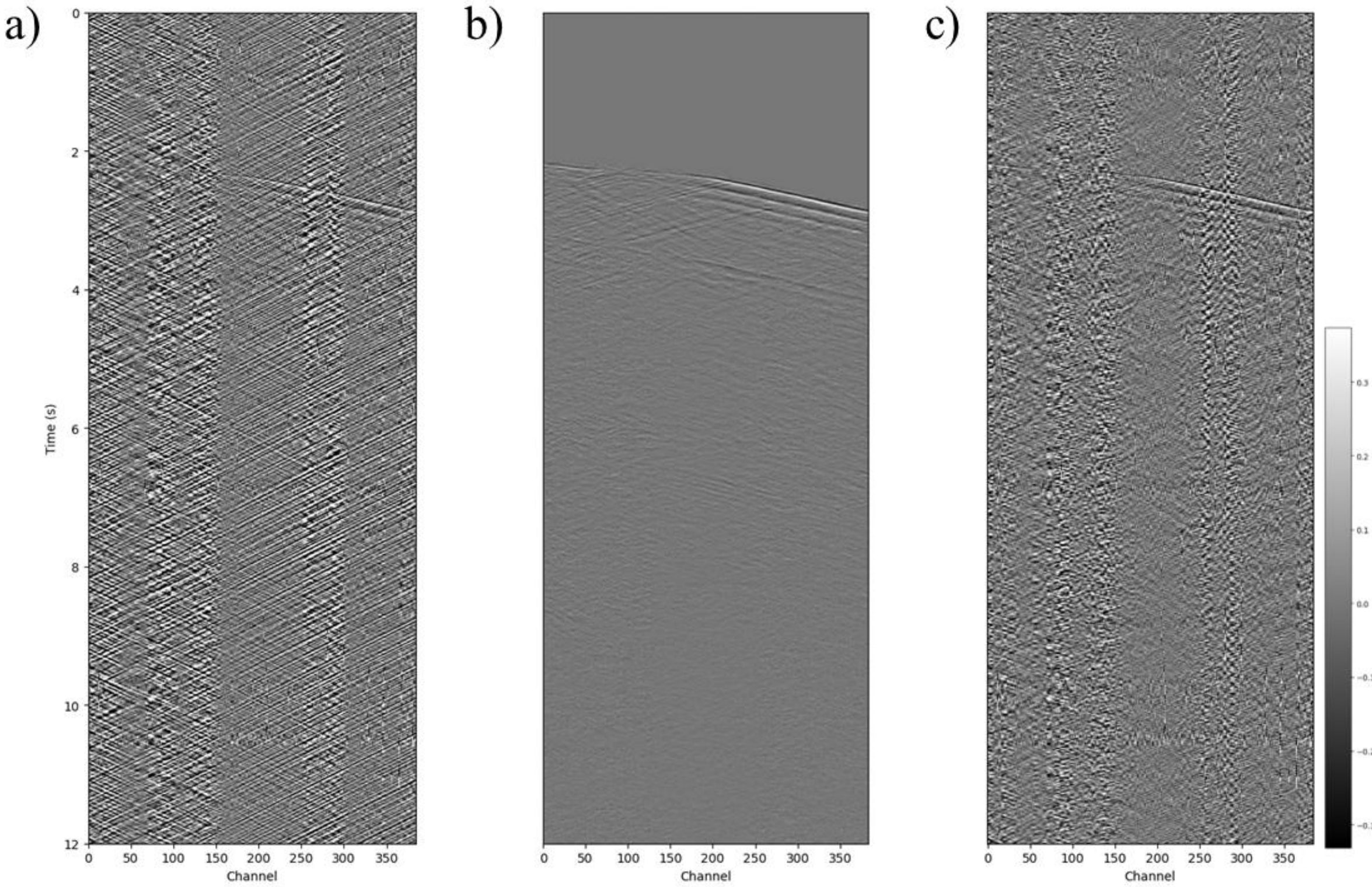

Figure 8. The denoising results on the field VSP record, a) raw data, b) denoising result from SeisDiff-denoNIA, c) FK filtering result.

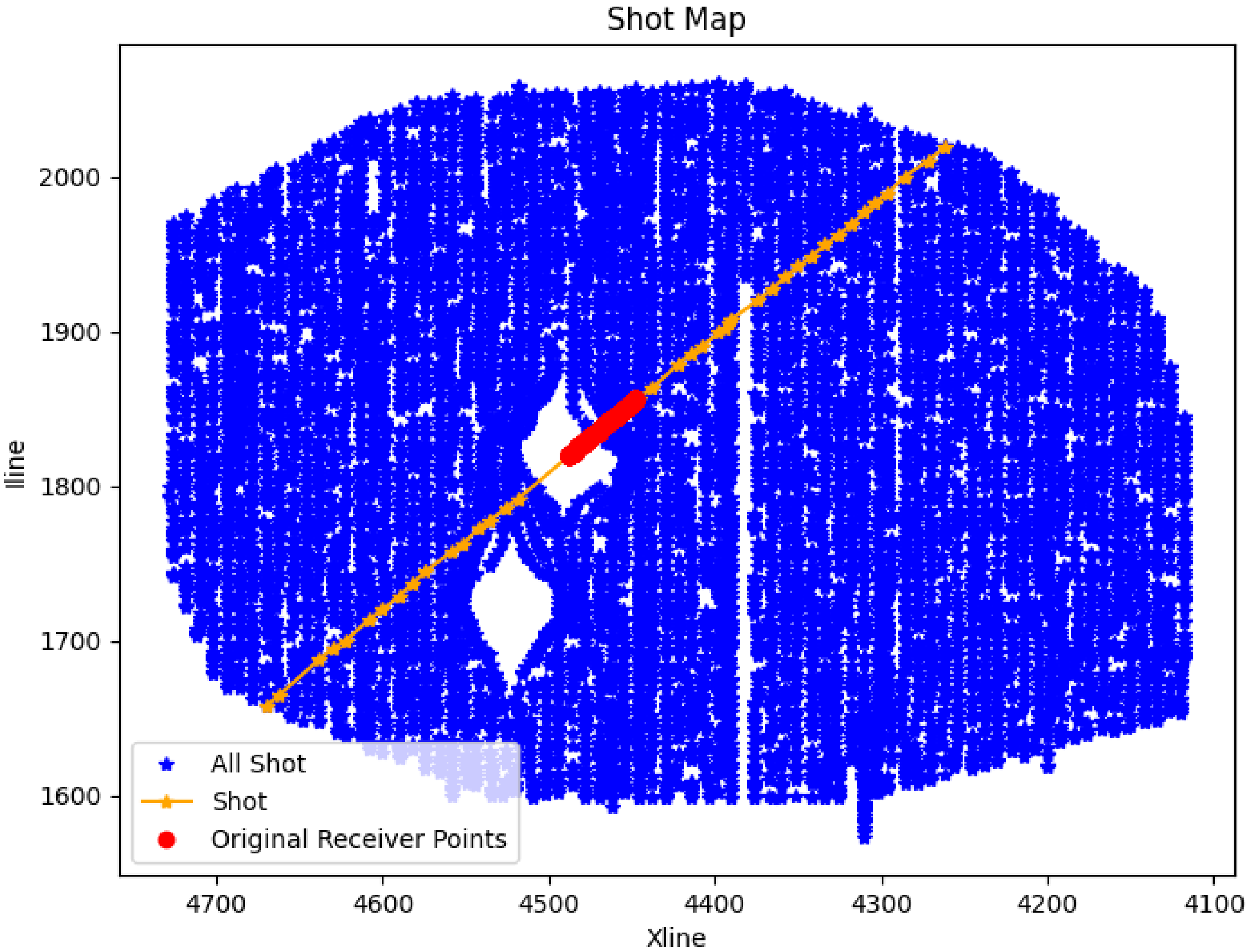

Figure 9. Shot map for the entire survey. Blue stars indicate the shot locations, red dots mark the surface positions corresponding to the receivers in the well, and orange markers denote the selected shots that are aligned with the receiver line on the surface.

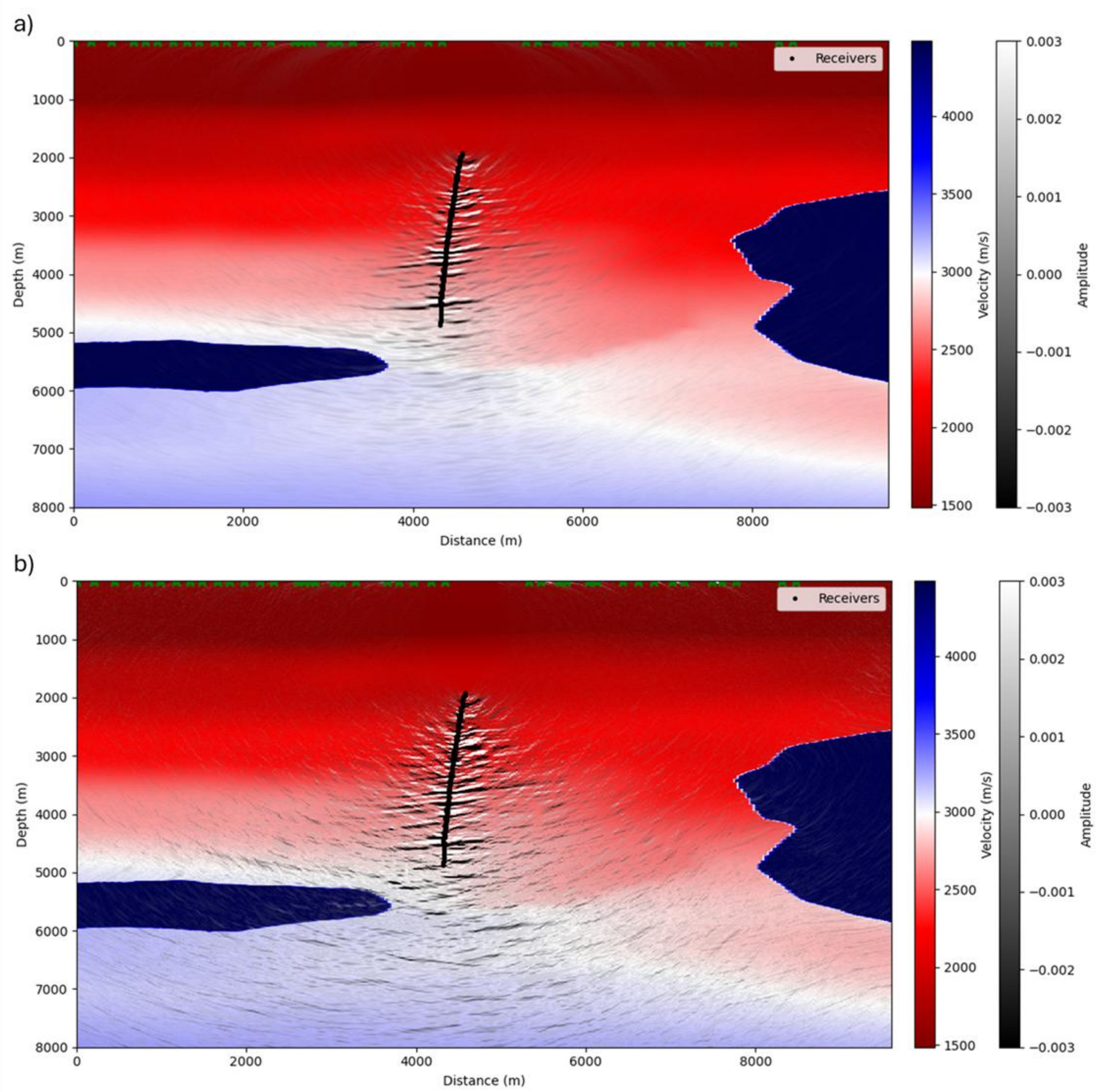

Figure 10. Pre-stack Kirchhoff imaging overlay with P-wave velocity model a) with proposed method denoising shots, b) with FK filtering shots. The green stars on the top represent shot locations.

## DISSCUSSION

### Ablation test on synthetic data

To quantitatively assess the advantages and robustness of the proposed method, we designed several controlled synthetic tests. We utilized the elastic VSP data from the SEAM Phase I model (Fehler and Larner, 2008) our clean reference dataset.

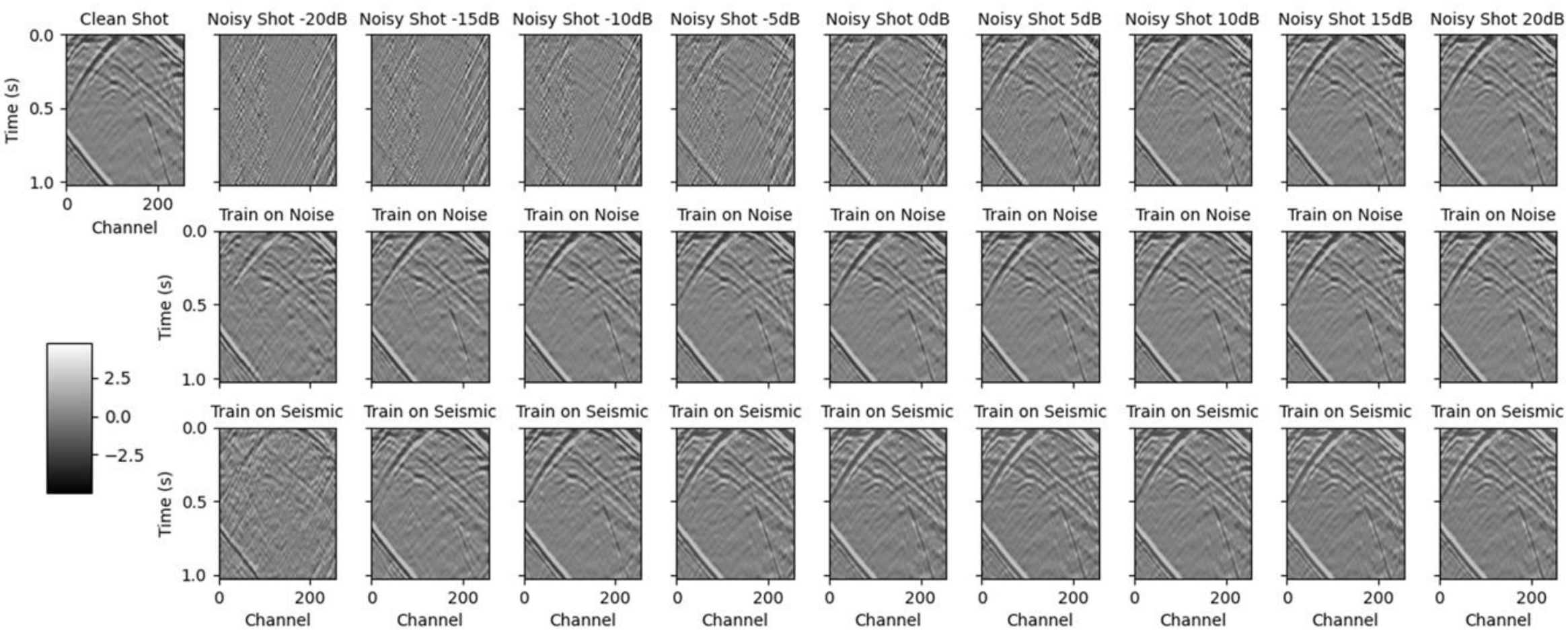

Figure 11. Synthetic test on SEAM VSP data with different SNR. The first row contains the clean shot patch and noisy shot patch in different noise levels. The second row contains the denoised results from the Train-on-Noise model. The last row contains the denoised results from the Train-on-Seismic model

Field noise extracted using our proposed workflow is added to the clean synthetic data at different SNR levels ranging from -20 dB to 20 dB. Two distinct diffusion models were initially trained to demonstrate the superiority of explicitly modeling noise: Train-on-Noise Model which is our proposed model, trained directly on extracted real field noise; Train-on-Seismic Model which is the conventional baseline diffusion model, trained using clean synthetic seismic data as the target labels—a common practice in seismic denoising literature. Figure 11 shows qualitative comparisons between clean data, noisy input, and denoised outputs at various SNR levels. At high SNRs, both models perform comparably, effectively removing noise and restoring the signal. However, in low SNR scenarios (in Figure 12), the Train-on-Noise model clearly outperforms its baseline counterpart by preserving more reflection details and signal continuity. Figure 13 presents the quantitative metrics, CC and RMSE, across varying SNR levels. The CC curve shows that structural similarity between clean and denoised data in the Train-on-Seismic model drops significantly below 0 dB SNR, whereas the Train-on-Noise model maintains high similarity down to -10 dB. RMSE values further support this observation, with the Train-on-Noise model achieving consistently lower amplitude errors, especially in noisy conditions.

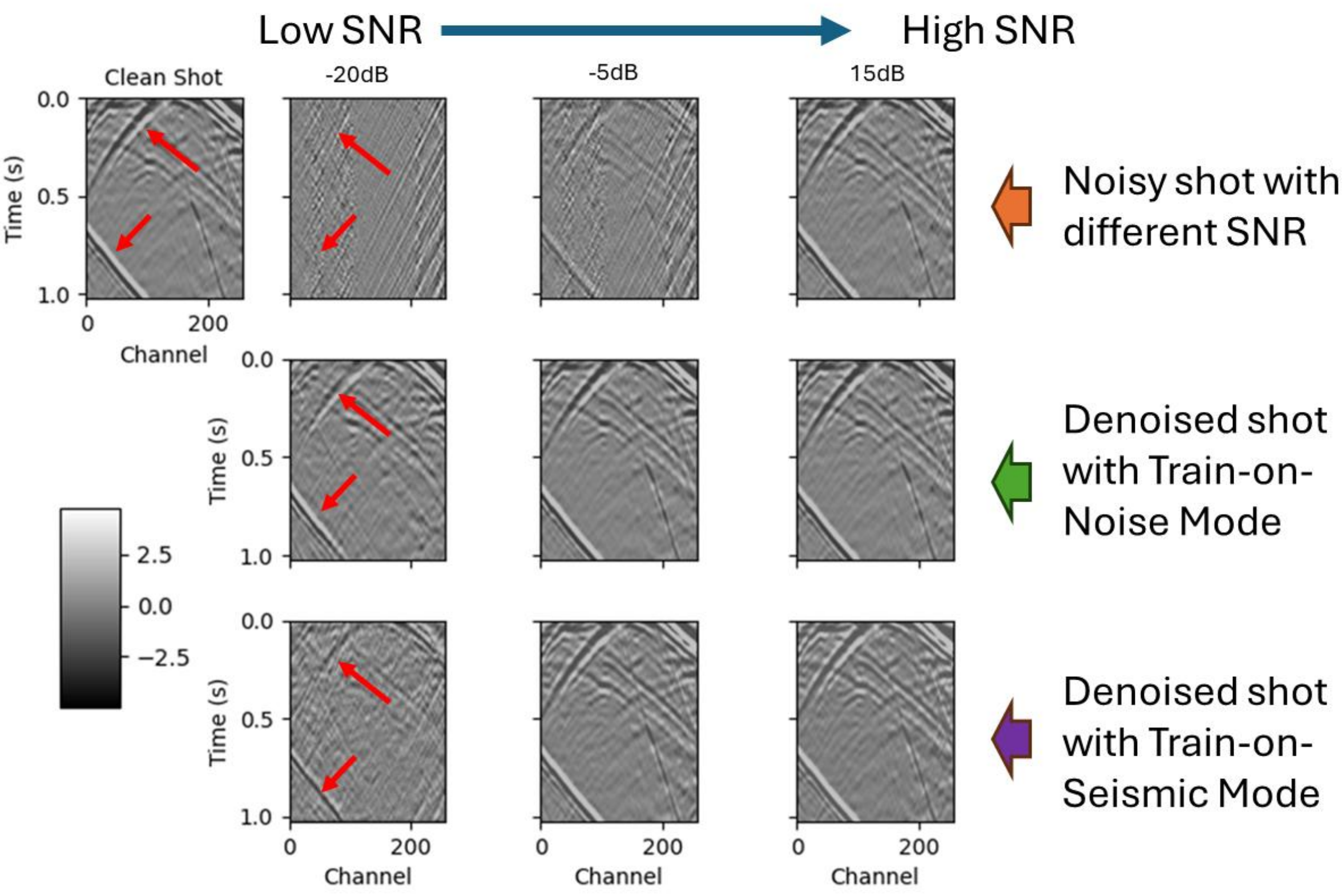

Figure 12. Selected results from Figure 11. The red arrows indicate the difference between denoised results and clean data for main reflectors.

**Ablation test on different perturbations**

This ablation study highlights the critical advantage of explicitly modeling noise distributions: the Train-on-Noise approach exhibits better generalization and robustness, particularly in challenging low SNR environments.

To further validate the robustness and targeted learning of our model towards noise characteristics rather than specific signal structures, we designed an additional ablation experiment. In this test, rather than adding synthetic seismic signals to the noise, we introduced simple hyperbolic events as artificial perturbations directly into the extracted noise samples (Figure 14). These hyperbolic perturbations simulate basic seismic-like events but do not resemble realistic seismic waveforms in complexity. The purpose was to determine whether the model's training targets the intrinsic noise patterns independently of the seismic signal structure.

The artificially perturbed noise images were subsequently used as conditional inputs to train another diffusion model under identical procedure as the Train-on-Noise model. We again quantified the denoising performance using the CC value to assess structural similarity, and the RMSE to measure amplitude fidelity (Table 2).

Interestingly, the Hyperbola-Perturbed Train-on-Noise model displayed only marginally reduced performance compared to the Seismic-Perturbed Train-on-Noise scenario, but obtained a comparable result to the Train-on-Seismic model. Although slightly lower CC values and higher RMSE values were observed, the differences were minor, indicating that the model predominantly focuses on learning intrinsic noise distributions rather than specific seismic structures. This slight degradation could be attributed to the hyperbolic perturbations being less representative of real seismic signals, causing mild ambiguity during the training of the conditional noise patterns.

Overall, this additional experiment supports our central argument: explicitly modeling the noise as the training target enhances model robustness and generalization. It also indicates the importance of the perturbation structure used during training, as overly simplistic or unrealistic perturbations may slightly degrade model performance compared to more realistic seismic perturbations. Nonetheless, the robustness and consistent performance of our approach remain clearly demonstrated, affirming the effectiveness of explicitly targeting real noise distributions.

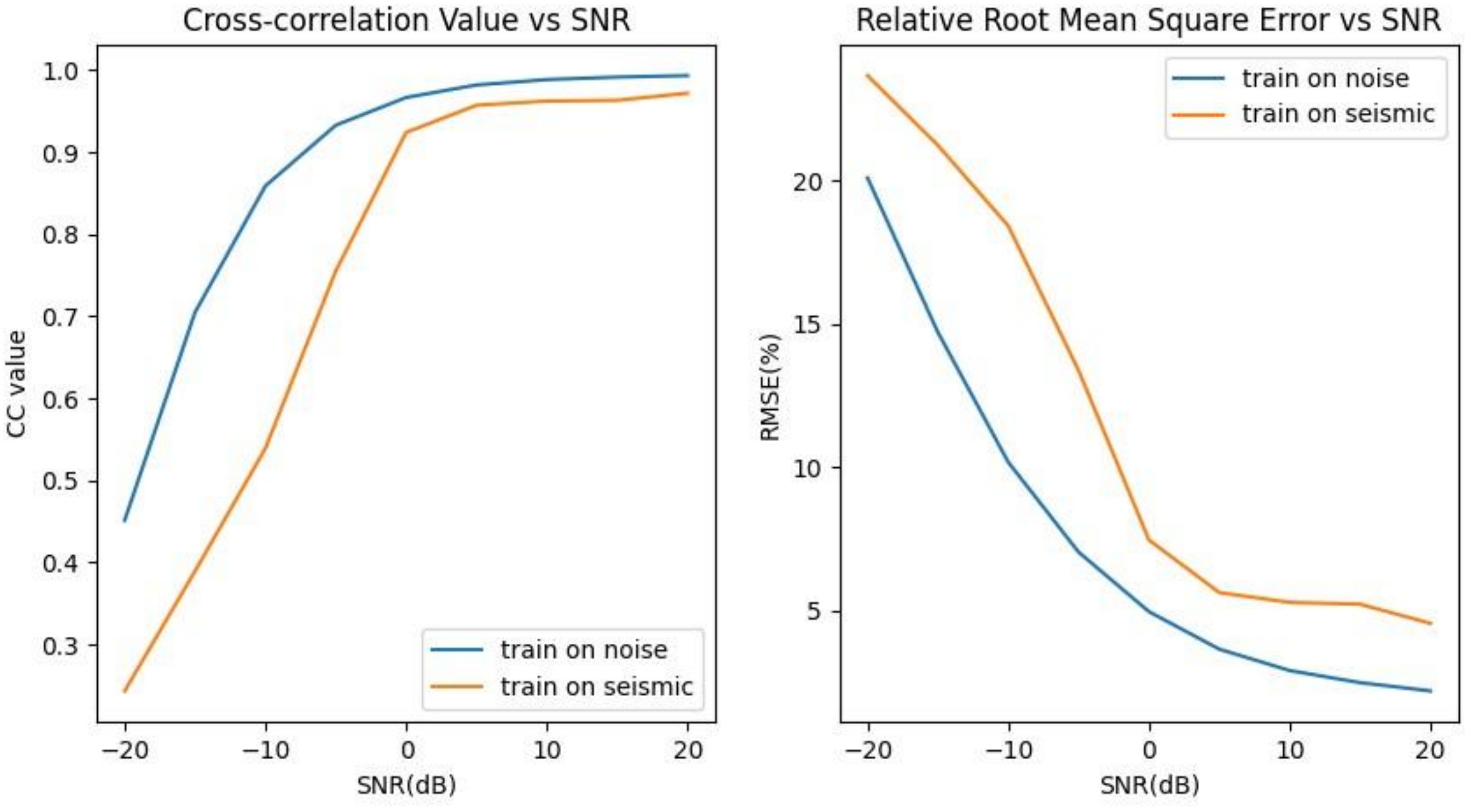

Figure 13. Quantitative evaluation of synthetic data test between Train-on-Noise and Train-on-Seismic

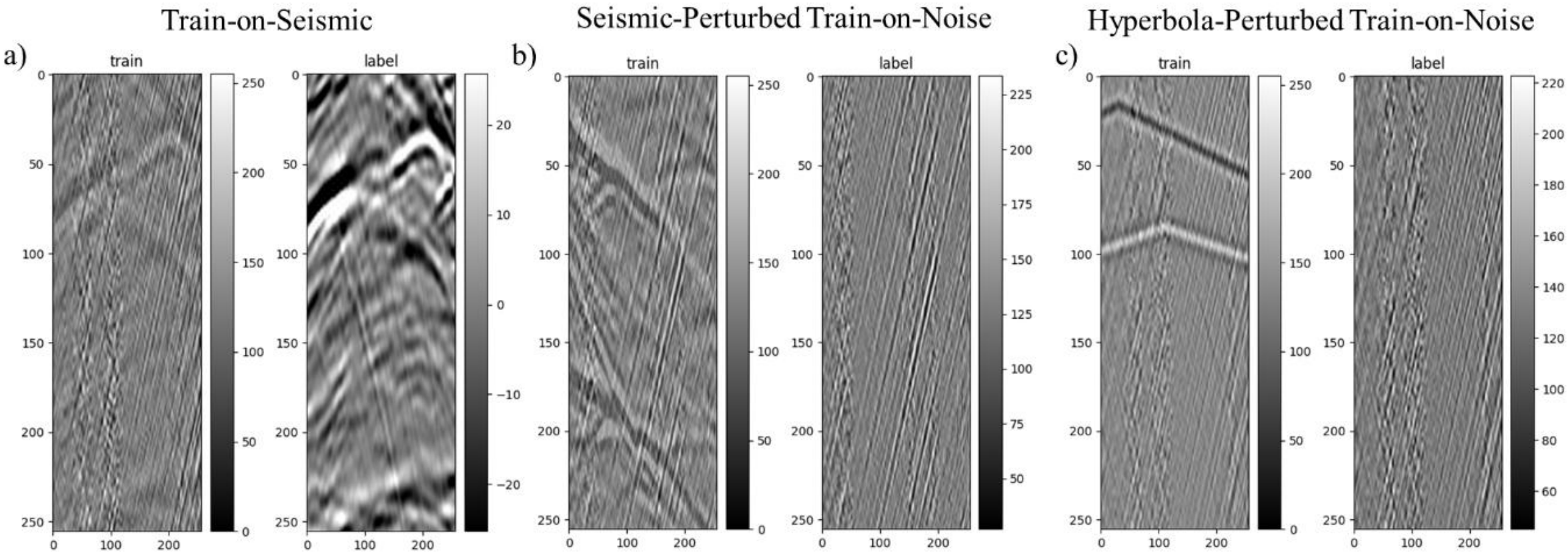

Figure 14. Examples for different training modes. a) training data pair for Train-on-Seismic mode, b) training data pair for Train-on-Noise mode with seismic signal as perturbation, c) training data pair for Train-on-Noise mode with hyperbola as perturbation.

Table 2. Quantitative evaluation of synthetic data test for different models. a) is CC value, b) is RMSE (%).

a)

| Model \ SNR | -20db | -10db | 0db | 10db | 20db |
|---|---|---|---|---|---|
| Train-on-Seismic | 0.24 | 0.54 | 0.92 | 0.96 | 0.97 |
| Hyperbola-Perturbed Train-on-Noise | 0.22 | 0.62 | 0.86 | 0.93 | 0.95 |
| Seismic-Perturbed Train-on-Noise | 0.45 | 0.86 | 0.97 | 0.99 | 0.99 |

b)

| Model \ SNR | -20db | -10db | 0db | 10db | 20db |
|---|---|---|---|---|---|
| Train-on-Seismic | 23.7 | 18.4 | 7.5 | 5.3 | 4.5 |
| Hyperbola-Perturbed Train-on-Noise | 27.0 | 16.3 | 10.1 | 7.2 | 6.2 |
| Seismic-Perturbed Train-on-Noise | 20.1 | 10.2 | 5.0 | 2.9 | 2.2 |

# CONCLUSION

We presented SeisDiff-denoNIA, a diffusion-based denoising framework that leverages true field noise as its primary learning target, eliminating reliance on synthetic clean–noisy training pairs. By explicitly modeling the noise distribution present in field DAS-VSP data, the method adapts naturally to diverse and challenging noise environments, including tube waves, fading, coupling-related noise, and production-induced disturbances. Our synthetic and field experiments demonstrate that this noise-driven strategy not only improves robustness under low-SNR conditions but also preserves key seismic events more faithfully than conventional signal-based diffusion models.

The successful application of SeisDiff-denoNIA to pre-stack migration further illustrates its practical value in real seismic processing workflows, enhancing reflection continuity and improving interpretability without introducing artifacts. These results point toward a broader paradigm shift in seismic denoising: rather than removing noise from the training process, we can use noise itself as a reliable, information-rich supervisory signal. Our results pave the way for a new class of synthetic-label-free, field-adaptive seismic processing techniques.

## ACKNOWLEDGEMENTS

This work was supported by the Reservoir Characterization Project at the Colorado School of Mines. The authors sincerely thank Shell for providing the datasets and for their support of this research.

REFERENCE

Abma, R., and J. Claerbout, 1995, Lateral prediction for noise attenuation by t-x and f-x techniques: Geophysics, **60**, 1887–1896.

Bekara, M., and M. van der Baan, 2009, Random and coherent noise attenuation by empirical mode decomposition: Geophysics, **74**, V89–V98.

Deighan, A. J., and D. R. Watts, 1997, Ground-roll suppression using the wavelet transform: Geophysics, **62**, 1896–1903.

Dong, X., Y. Li, T. Zhong, N. Wu, and H. Wang, 2022, Random and Coherent Noise Suppression in DAS-VSP Data by Using a Supervised Deep Learning Method: IEEE Geoscience and Remote Sensing Letters, **19**, 1–5.

Durall, R., A. Ghanim, M. Fernandez, N. Ettrich, and J. Keuper, 2023, Deep Diffusion Models for Seismic Processing: Computers & Geosciences, 17, August 2023, 105377.

Fehler, M., and K. Larner, 2008, SEG Advanced Modeling (SEAM): Phase I first year update: Leading Edge, **27**, 1006–1007.

Ho, J., A. Jain, and P. Abbeel, 2020, Denoising Diffusion Probabilistic Models: ArXiv,

https://doi.org/10.48550/arXiv.2006.11239, accessed Dec 2025.

Konietzny, S., V. H. Lai, M. S. Miller, J. Townend, and S. Harmeling, 2024, Unsupervised coherent noise removal from seismological Distributed Acoustic Sensing data: Journal of Geophysical Research: Machine Learning and Computation, **1**, e2024JH000356.

Li, Y., H. Zhang, J. Huang, and Z. Li, 2024, Conditional denoising diffusion probabilistic model for ground-roll attenuation: ArXiv, https://doi.org/10.48550/arXiv.2403.18224, accessed Dec 2025.

Liu, S., C. Birnie, and T. Alkhalifah, 2023, Trace-wise coherent noise suppression via a self-supervised blind-trace deep-learning scheme: Geophysics, **88**, V459–V472.

Lu, C., Y. Zhou, F. Bao, J. Chen, C. Li, and J. Zhu, 2022a, DPM-Solver: A Fast ODE Solver for Diffusion Probabilistic Model Sampling in Around 10 Steps: ArXiv, https://doi.org/10.48550/arXiv.2206.00927, accessed Dec 2025.

Lu, C., Y. Zhou, F. Bao, J. Chen, C. Li, and J. Zhu, 2022b, DPM-Solver++: Fast Solver for Guided Sampling of Diffusion Probabilistic Models: ArXiv, https://doi.org/10.48550/arXiv.2211.01095, accessed Dec 2025.

Luiken, N., M. Ravasi, and C. Birnie, 2024, Integrating self-supervised denoising in inversion-based seismic deblending: Geophysics, **89**, WA39–WA51.

Saad, O. M., and Y. Chen, 2020, Deep denoising autoencoder for seismic random noise attenuation: Geophysics, **85**, V367–V376.

Wang, X., S. Fan, C. Zhao, D. Liu, and W. Chen, 2023, A self-supervised method using Noise2Noise strategy for denoising CRP gathers: IEEE Geoscience and Remote Sensing Letters : A Publication of the IEEE Geoscience and Remote Sensing Society, **20**, 1–5.

Yang, L., S. Fomel, S. Wang, X. Chen, W. Chen, O. M. Saad, and Y. Chen, 2023, Denoising of distributed acoustic sensing data using supervised deep learning: Geophysics, **88**, WA91–WA104.

Zheng, K., C. Lu, J. Chen, and J. Zhu, 2023, DPM-solver-v3: Improved diffusion ODE solver with empirical model statistics: ArXiv, https://doi.org/10.48550/arXiv.2310.13268, accessed Dec 2025.

Zhu, D., P. Li, and G. Jin, 2025, SeisDiff-deno: A diffusion-based denoising framework for tube wave attenuation in VSP data: ArXiv, https://doi.org/10.48550/arXiv.2503.00637, accessed Dec 2025.

Zhu, D., L. Fu, V. Kazei, and W. Li, 2023, Diffusion Model for DAS-VSP Data Denoising: Sensors , **23**, 8619.

Zwartjes, P., A. Mateeva, D. Chalenski, Y. Duan, D. Kiyashchenko, and J. Lopez, 2018, Frequent, multiwell, stand-alone 3D-DAS VSP for low-cost reservoir monitoring in deepwater: SEG Technical Program Expanded Abstracts 2018.